\documentclass[a4paper,twoside]{article}
\newcommand{\affil}[1]{$^{\rm #1}$}
\usepackage[authoryear]{natbib}
\bibpunct{(}{)}{;}{a}{}{,}
\usepackage{graphicx}
\date{} 
\newcommand{\OIII}{\mbox{[O\,{\sc iii}]}}
\newcommand{\NII}{\mbox{[N\,{\sc ii}]}}
\title{\large\bf\flushleft Sub-arcsecond Morphology of Planetary Nebulae}
\author{\parbox{\textwidth}{\flushleft
\vspace{-0.5cm}
{\it L.~F. Miranda\affil{A,B,D}, G. Ramos-Larios\affil{A,C}, and M.~A. Guerrero\affil{A}}\\
\vspace{0.4cm}
{\small \affil{A}\,Instituto de Astrof\'{\i}sica de Andaluc\'{\i}a - CSIC, Granada (Spain)}\\
{\small \affil{B}\,Facultade de Ciencias, Universidade de Vigo, Vigo (Spain)
  [Current address] } \\
{\small \affil{C}\,Instituto de Astronom\'{\i}a y Meteorolog\'{\i}a, Guadalajara (Mexico)}\\
{\small \affil{D}\,Email: lfm@iaa.es}}}
\begin{document}
\twocolumn[
\begin{changemargin}{.8cm}{.5cm}
\begin{minipage}{.9\textwidth}
\vspace{-1cm}
\maketitle
%
%
\small{\bf Abstract: Planetary nebulae (PNe) can be roughly categorized into several broad 
morphological classes. The high quality images of PNe acquired in recent 
years, however, have revealed a wealth of fine structures that preclude simplistic models 
for their formation. Here we present narrow-band, sub-arcsecond images of 
a sample of relatively large PNe that illustrate the complexity and variety 
of small-scale structures. This is especially true for bipolar PNe, for 
which the images reveal multi-polar ejections and, in some cases, suggest
turbulent gas motions. Our images also reveal the presence or signs of
jet-like outflows in several objects in which this kind of component has not
been previously reported.}


\medskip
\medskip
\end{minipage}
\end{changemargin}
]
\small

\section{Introduction}

Planetary nebulae (PNe) consist of material ejected from the
Asymptotic Giant Branch (AGB) progenitor, which is photoionized by the
hot central star and swept up by the fast stellar wind
\citep{KOW78}. The morphology of PNe has attracted the attention
of many observers because it is related to and contains information
about the processes of mass ejection involved in their formation. 
A number of investigations have been devoted to the morphological
classification of PNe \citep{BAL87,SCH92,STA93,MAN96}.  In general,
three broad morphological classes have been considered: round (or
circular), elliptical and bipolar. The formation of the three classes
was explained within the Generalized Interacting Stellar Winds (GISW)
Model \citep{BAL87} assuming an azimuthal dependence of the density in the
slow wind, having its maximum at the equatorial plane. Bipolar and elliptical 
PNe would present ``high'' and ``intermediate'' density contrasts, while round PNe
would present a homogeneous slow wind. Numerical simulations have shown that
the three main morphological classes can be reproduced under this
hypothesis \citep[e.g.,][]{FRA93}.

The detection of collimated outflows in PNe
\citep[e.g.,][]{GIE85,MIR92} poses a serious difficulty to the GISW
model because this model is unable to explain the presence of highly
collimated ejections. Moreover, no collimating agent was foreseen in
these evolutionary phases, which could cause high collimation. In
addition, the collimated ejections in most PNe present different
orientations with respect to the central star suggesting that the
collimated agent can precess or wobble \citep{MIR92,MIR99,GUE08}. 

More recently, high resolution images obtained with the {\it HST} have
shown that PNe are highly complex objects and that simplistic
models are unable to account for the large variety of structures and
microstructures observed in these objects. In particular, the
signature of collimated outflows appears in a very large fraction of
PNe \citep[e.g.,][]{SAH98}. Many PNe show multiple bipolar ejections with
different geometries that 
require episodic ejections \citep[e.g.,][]{GUE04}. These observations have led 
to suggest that collimated outflows are the basic mechanism that shapes 
PNe \citep{SAH98}.  Nevertheless, other physical phenomena should be also 
present in PN formation as wind interaction, magnetic fields and/or the binary nature of the central
star \citep[see][]{BYF02}. Although a large number of PNe have already been
imaged \citep{BAL87,SCH92,MAN96,SAH98}, in some cases the images lack spatial resolution and/or 
quality to enable a proper identification of the morphological components. In fact,
images at higher resolution and/or deeper than previous ones continue
discovering new structures, even in well observed PNe, which result to be
key for a proper interpretation of their formation
\citep[e.g.,][]{MGM04,MIR06}. These results and the complexity of PNe 
clearly justify the imaging of PNe whose images are of insufficient quality.

In this paper, we present narrow-band images of a sample of relatively 
large PNe, including elliptical and bipolar PNe. The images have been 
taken under particularly good seeing conditions and reveal novel
details in the objects.

\begin{figure}[h]
\begin{center}
\includegraphics[scale=0.8, angle=0]{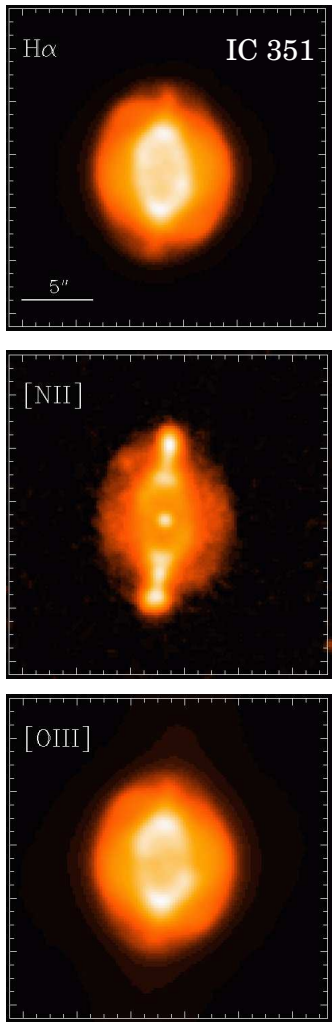}
\caption{H$\alpha$, {\OIII} and {\NII} images of IC\,351. North is up, east to the left. 
The images are displayed in a logarithmic scale. White represents high values of the intensity.}\label{fig1}
\end{center}
\end{figure}

\begin{figure}[h]
\begin{center}
\includegraphics[scale=0.8, angle=0]{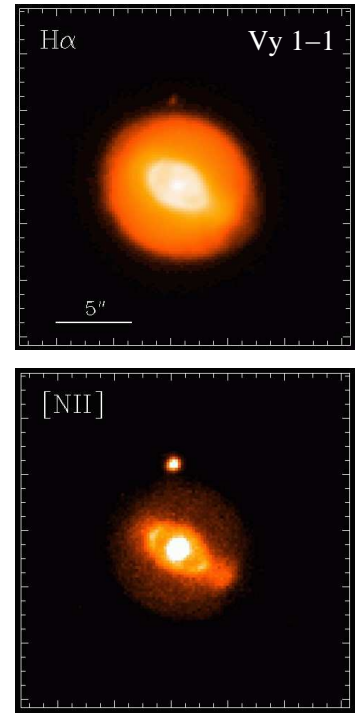}
\caption{
H$\alpha$ and {\NII} images of Vy\,1-1. North is up,
east to the left.  The images are displayed in a logarithmic scale. White represents high values of the
intensity.}\label{fig2}
\end{center}
\end{figure}

\section{Sample selection and observations}

The observed PNe were selected from the \citet{MAN96} catalog among
those with no independent H$\alpha$ and {\NII} images and/or with low
resolution images due to poor seeing conditions. Images isolating the
H$\alpha$ and {\NII} lines are crucial because these
two emission lines may trace very different regions of the
nebulae; in particular, collimated outflows and microstructures can be
better recognized and studied in {\NII}. In addition,
good seeing conditions allow us a detailed morphological study of the
structures present in a PN. We have added IC\,5217 and NGC\,6778 to this
sample in order to study at high spatial resolution the structures
identified by \citet{MIR06} and \citet{MGM04}, respectively.

Direct images were obtained on 2006 June and 2008 September using
ALFOSC at the 2.56\,m Nordic Optical Telescope on Roque de los
Muchachos Observatory (La Palma, Spain). A 2K$\times$2K E2V CCD 
with plate scale of 0.19$''$\,pixel$^{-1}$ was used as detector. We used three narrow-band
filters that isolate the light of H$\alpha$ ($\Delta$$\lambda$ = 9 {\AA}), [N\,{\sc
ii}]$\lambda$6584 (hereafter {\NII}) ($\Delta$$\lambda$ = 9 {\AA}), and [O\,{\sc iii}]$\lambda$5007
(hereafter {\OIII}) ($\Delta$$\lambda$ = 30 {\AA}). Exposure time per filter was
900\,s or 1800\,s. The seeing during the observations was between 0.5$''$ and
0.9$''$. The images were reduced using standard IRAF routines.

\section{Results and Discussion}

In the following we will discuss the images of the objects
observed. In some cases, we have grouped together some objects
because of their morphological similarities.

\subsection{IC\,351 and Vy\,1-1}

In the \citet{MAN96} catalog, IC\,351 is classified as an elliptical PNe with
internal structures while Vy\,1-1 is classified as an elliptical
multiple-shell PN. The two objects are not included in the lists of PNe with low-ionization
structures, knots or jets by \citet{DEN01}

Figures\,1 and 2 show our images of IC\,351 and Vy\,1-1,
respectively. The two PNe present noticeable morphological 
similarities, including low-ionization polar structures 
which are identified in these images for the first time. 

IC\,351 is a high-excitation PN consisting of an inner elliptical shell oriented along PA $\sim$
355$^{\circ}$ surrounded by an outer round attached shell. The central star is detected
in the {\NII} and {\OIII} filters.  The {\NII} image shows
two collimated structures emanating from bright polar caps of the
inner elliptical shell that extend $\sim$ 3$''$ and end in bright knots. The H$\alpha$ and {\OIII}
images also show hints of these structures but they are relatively
much fainter, implying low-excitation.

Vy\,1-1 (Figure\,2) shows an inner elliptical shell oriented at PA
$\sim$ 70$^{\circ}$, and an outer round attached shell. Although
no {\OIII} image has been obtained for Vy\,1-1, the faint emission in {\NII}
and the bright emission in {\OIII} \citep[see][]{MAN96} indicate that both shells
are of high-excitation. The polar regions of the inner shell are particularly bright in
{\NII}. Two polar knots are observed outside but connected to the
elliptical inner shell in the {\NII} image; they are not recognizable in the
H$\alpha$ image. The south-western knot is brighter than the
north-eastern one.

\begin{figure}[h]
\begin{center}
\includegraphics[scale=0.7, angle=0]{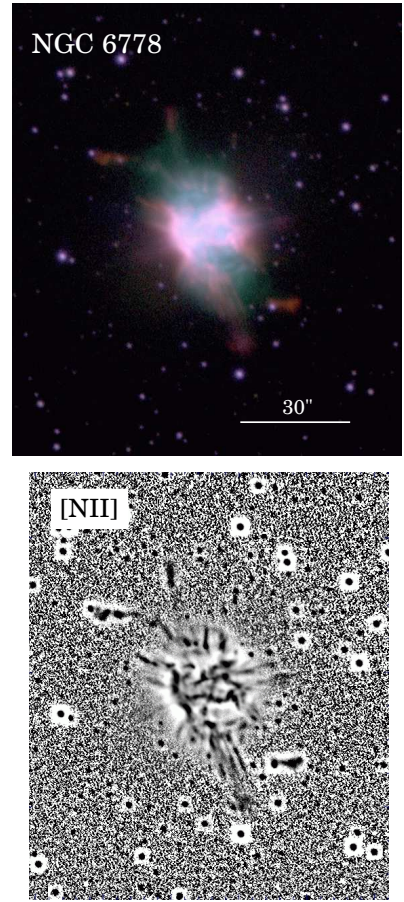}
\caption{
(top) Colour composite picture of NGC\,6778 (green = H$\alpha$, blue =
{\OIII} and red = {\NII}).  The image is displayed in a logarithmic
scale. (bottom) Unsharp masking {\NII} image. The image is 
displayed in a linear scale. North is up, east
to the left in the images.}\label{fig3}
\end{center}
\end{figure}

\begin{figure}[h]
\begin{center}
\includegraphics[scale=0.6, angle=0]{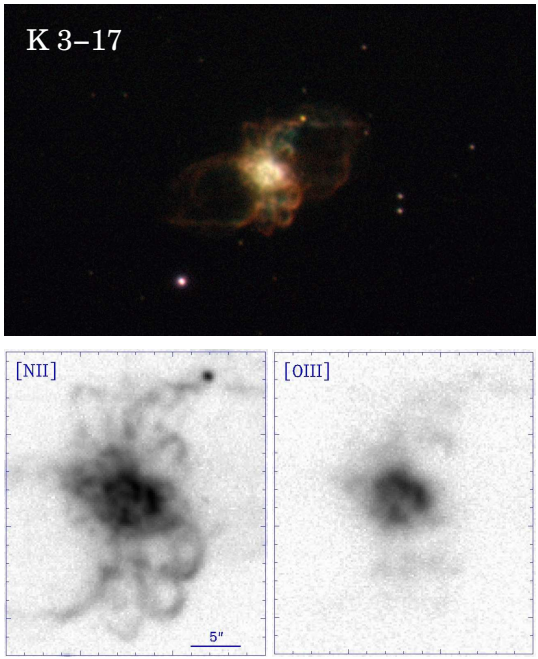}
\caption{
(top) Colour composite picture of K\,3-17 (green = H$\alpha$, blue =
{\OIII} and red = {\NII}). (bottom) Enlargement of the bright central
region in the {\NII} and {\OIII} images. The images are displayed in a
logarithmic scale. North is up, east to
the left in the images.  The spatial scale is indicated in the {\NII}
image.}\label{fig4}
\end{center}
\end{figure}

IC\,351 and Vy\,1-1 closely resemble NGC\,6826 and NGC\,7009
\citep{BAL98}. Although the radial velocities of
the polar structures in IC\,351 and Vy\,1-1 have not been measured, it
is probable that they represent FLIERs as in the case of
NGC\,6826 and NGC\,7009. The morphological resemblances
suggest similar formation processes in the four PNe.

\subsection{NGC\,6778}

NGC\,6778 is a PN that contains two bipolar jet systems oriented at 
different directions and moving at 100--200 km\,s$^{-1}$ \citep{MGM04}. 
Figure\,3 show a colour composite picture of NGC\,6778 constructed with the H$\alpha$, {\OIII} and
{\NII} images. The high resolution of the new images allow us a 
detailed description of the nebulae. The appearance of NGC\,6778 is
that of a bipolar nebula with its major axis along PA $\simeq$
15$^{\circ}$. The nebula does not display the characteristic waist of
butterfly PNe, but the region along its minor axis is defined by a
fragmented structure that is particularly bright in {\NII}. In the
high-excitation [O\,{\sc iii}] emission, the nebula appears elliptical rather
than bipolar. The images reveal details of the bipolar jets. In
particular, the jet oriented at PA $\simeq$ 195$^{\circ}$ consists of
at least four filaments emanating from a bright knot. In addition,
the nebula presents many low-excitation knots with a cometary
appearance, embedded in high-excitation material. These knots are more
clearly observed in the unsharp masking {\NII} image also shown in
Figure\,3. This image further strengthens the complexity of the nebula. We
found interesing similarities between 
NGC\,6778 and Sh\,2-71 \citep{BOH01,MIR05} as both PNe exhibit 
extremely knotty and filamentary morphologies that suggests that the previously existing 
structures have been fragmented and probably swept up by the fast stellar wind.

\subsection{K\,3-17}
\begin{figure}[h]
\begin{center}
\includegraphics[scale=0.5, angle=0]{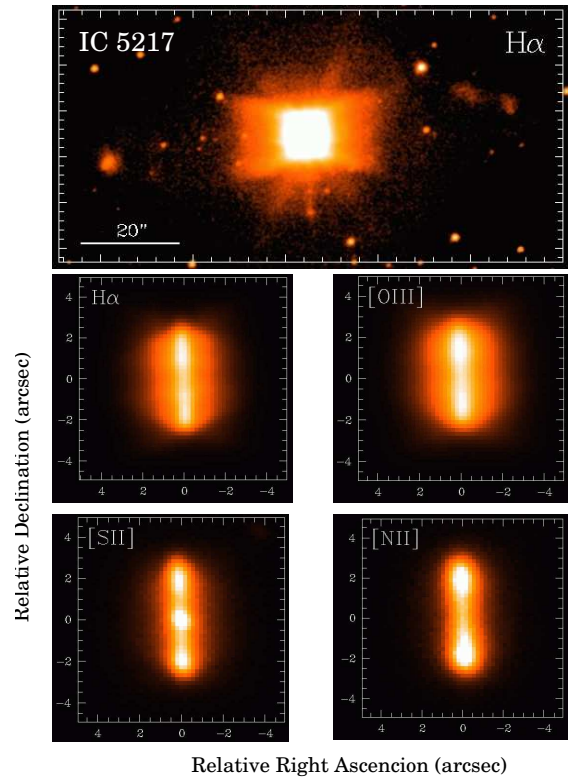}
\caption{
(top) H$\alpha$ image of IC\,5217 showing the whole bipolar structure
and the point-symmetric features. North is up, east to the left. (bottom) Images of the equatorial
regions in several filters (upper left). The central star is observed
in the [S\,{\sc ii}] filter. The images are displayed in a logarithmic scale.}\label{fig5}
\end{center}
\end{figure}

K\,3-17 presents a bipolar morphology in the \citet{MAN96} catalog
consisting of a bright compact core and two faint bipolar lobes in the
H$\alpha$+{\NII} image. In the {\OIII} image only the bright compact
core was detected. Our new, higher
resolution images reveal that K\,3-17 is a complex PN with a
wealth of structures. A colour composite picture is shown in Figure\,4
along with {\NII} and {\OIII} images of the bright core. In the new
images, the bipolar lobes present a spindle-like shape and are
dominated by {\NII} emission. The core is resolved into a series of
bubbles oriented mainly perpendicular to the bipolar lobes, although
small bubbles oriented along the bipolar axis are also observed. No
ring-like or toroidal structure can be identified in the center but
bright knots without a particular orientation. Most of the
structures are of low-excitation while high-excitation is only
observed at the central knots and in the regions of the lobes near the center.

The spindle-like morphology of the lobes suggests the action of
collimated outflows in their formation. A similar morphology is observed in 
Hu\,2-1 in which the distorsions of the lobes can be attributed due to the
action of a bipolar collimated 
outflows \citep{MIR01}. As for the origin of the
equatorial bubbles in K\,3-17, it could be possible that the stellar wind is
protruding through or breaking an original equatorial
structure. Alternatively, the bubbles may represent collimated
outflows along different directions. In this case, the orientation of the 
collimation axis should have changed drastically by $\simeq$
90$^{\circ}$. Large differences in the orientation of multipolar lobes are also
found in other PNe, being Sh\,2-71 an extreme case \citep{MIR05}, while the
equatorial bubbles in K\,3-17 are similar to these observed in {\it HST}
images of Hb\,5 \citep[see][]{MON09}.

\section{IC\,5217 and KjPn\,6}

IC\,5217 is an edge-on bipolar PN with polar point-symmetric features, a large 
axial ratio and a narrow waist that contains a very thin, bright equatorial
ring \citep{MIR06}. 

\begin{figure}[h]
\begin{center}
\includegraphics[scale=0.8, angle=0]{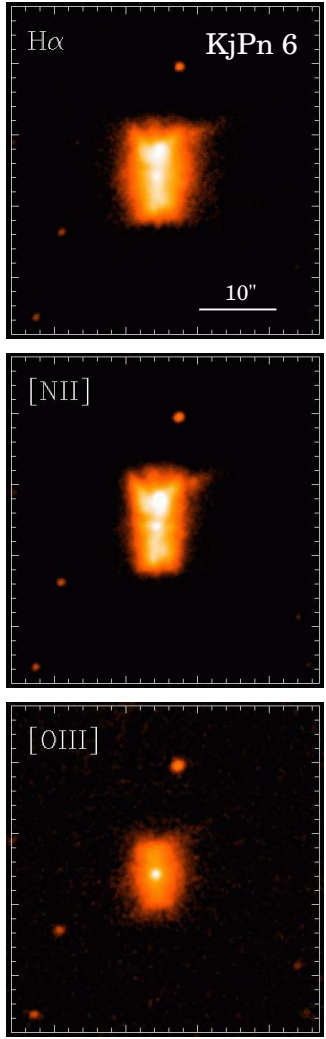}
\caption{H$\alpha$, {\NII}, and {\OIII} images of KjPn\,6.  The images
  are displayed in a logarithmic scale. North is up, east to left}\label{fig6}
\end{center}
\end{figure}

Figure\,5 presents a new H$\alpha$ image and images of the ring in several
emission lines. The large scale H$\alpha$ image
is similar to that presented by \citet{MIR06}. In particular, the point-symmetric 
structures at large distances from the center are observed in this image as
well as the bipolar lobes emanating from the nebular central regions that seem
to be connected with the point-symmetric
features. The higher resolution
of the new images and the observations in several emission line filters 
allow us to analyze the bright equatorial ring in great detail. Four local
maxima along the ring are distinguished in the
H$\alpha$ image in agreement with the radio continuum morphology at 3.6\,cm
\citep[see][]{MIR06}. In the rest of the emission lines, a clear
ionization gradient is observed with the radius of the ring in {\OIII} being 
smaller than in {\NII} and [S\,{\sc ii}]. We also note that 
the central star is detected in the [S\,{\sc ii}] image at the center of the ring. It
is worth noting that, if only a short exposure image would have been
acquired, IC\,5217 would be described  
as a highly collimated bipolar jet emanating from the central star. This is
not the case as high-resolution, long-slit spectroscopy 
demonstrates that the bright structure is an edge-on ring \citep{MIR06}.

KjPn\,6 presents a peculiar, almost triangular shape in the
H$\alpha$+{\NII} image and extremely faint emission in the {\OIII}
image \citep{MAN96}. The new images of KjPn\,6 are shown in
Figure\,6. The structure in H$\alpha$ and {\OIII} resembles an
elliptical PN with the major axis near the north-south direction,
while in {\NII} the nebula appears more squarish. Faint details are
observed in the inner regions, which are particularly bright in {\NII}
and H$\alpha$ but absent in {\OIII}. The morphology of KjPn\,6 is very similar
to this observed in the inner regions of IC\,5217 (Figure\,5), which have been
identified with an edge-on ring. These similarities lead us to suggest that KjPn\,6
may be a bipolar PN with bipolar much fainter than its equatorial
regions. Therefore, IC\,5217 and KjPn\,6 are probably related to ring-like PNe
that are characterized by the 
presence of a bright ring accompanied by faint bipolar lobes, although these are not 
always detected \citep[e.g.,][]{BPW03}. Interesting similarites are found with
the edge-on ring-like PNe Me\,1-1 \citep{PER08} and IC\,2149
\citep{VAZ02}.

\section{Bipolar planetary nebulae}

Figure\,7 shows colour composite pictures of six bipolar PNe. The new images are 
deeper and of higher quality than these published elsewhere. BV\,1 is an
edge-on bipolar with a bright ring and extremely faint bipolar
lobes \citep{KAL88}. HaTr\,10, M\,4-17, and K\,3-46 are likely similar
to BV\,1, the main difference being the orientation
of the polar axis with respect to the observer and/or the evolutionary
stage. The case of NGC\,650 could be different as this PN 
is observed almost edge-on, as is the case of BV\,1, but the equatorial region
is thick and there is not a large intensity contrast between this equatorial
region and the bipolar lobes. In addition, NGC\,650 shows bubbles and microstructures that
suggest the presence of focused (collimated) outflows. M\,2-48 is a
bipolar PN with multiple structures and high velocity collimated
outflows \citep{VAZ00,LOP02}. Our high resolution image (Figure\,7) shows in great 
detail the structures previously detected. A dark lane, rather than a 
bright equatorial torus or ring, separates the bipolar lobes. Elongated protrusions are 
observed along the minor axis of the nebula, although their orientation is
different from this of the bipolar lobes and collimated outflows. 
A series of knots encircle the bipolar lobes tracing an apparent circular
shell. This shell is brighter at four regions along the main
symmetry axis and its center does not coincide with the nebular
core. The distant outflows 
discovered by \citet{VAZ00} appear as knotty structures, particularly the
north-eastern one. Its south-western counterpart is clearly detected in the new
images, while only hints of it were found by
\citet{VAZ00}. M\,2-48 presents noticeable 
differences when compared to the other bipolar PNe in Figure\,7. On the other
hand, the presence of distant outflows in M\,2-48 resembles the situation
observed in IC\,5217 (see above).  

\begin{figure*}[h]
\begin{center}
\includegraphics[scale=1.2, angle=0]{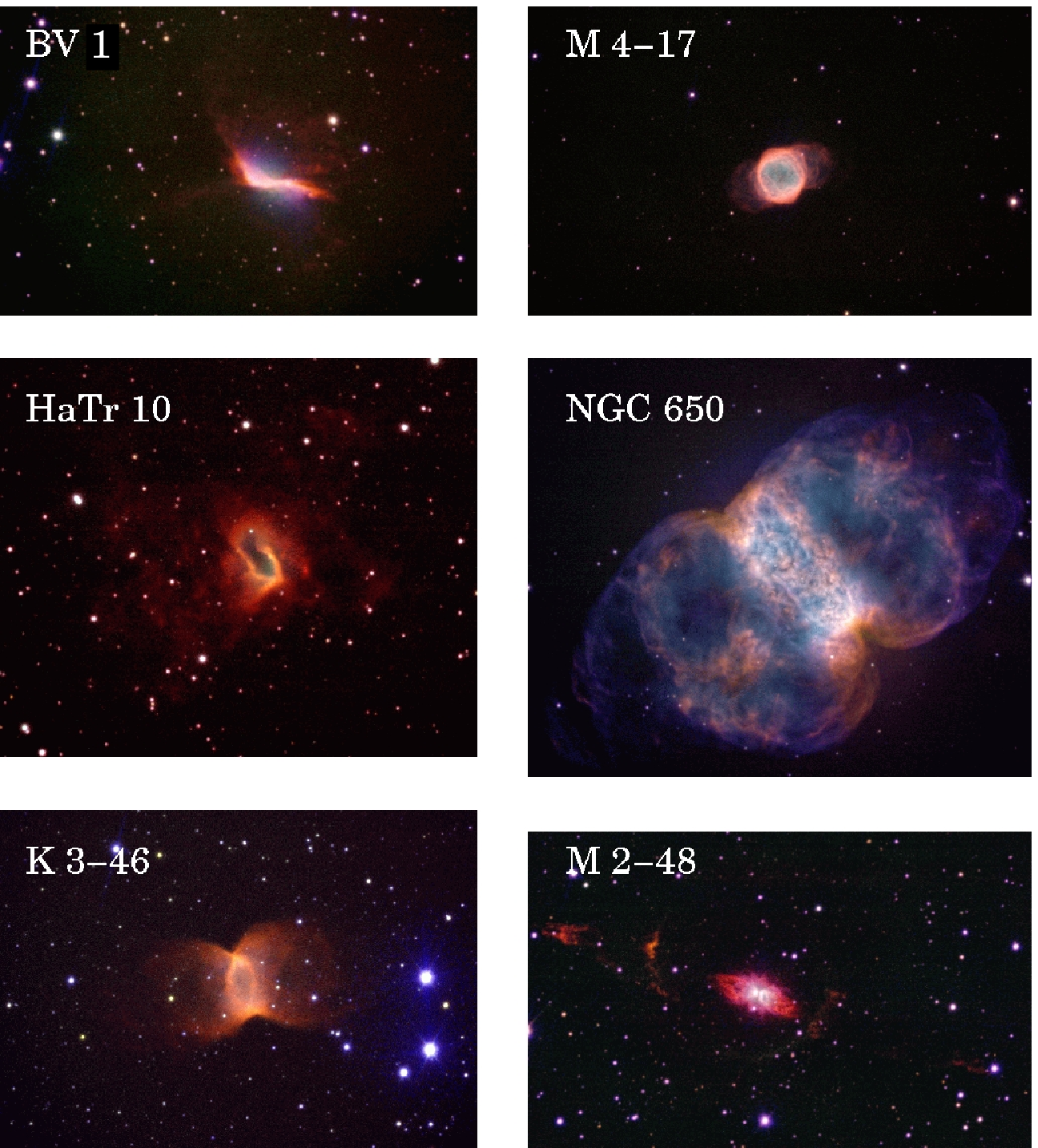}
\caption{Colour composite pictures of BV\,1 (field of view [fov] $\simeq$
  285$''$$\times$155$''$), 
M\,4-17 (fov $\simeq$ 255$''$$\times$130$''$),  HaTr\,10 (fov $\simeq$
170$''$$\times$140$''$), NGC\,650 (fov $\simeq$ 130$''$$\times$105$'$), 
K\,3-46 (fov $\simeq$ 215$''$$\times$150$''$), and M\,2-48 (fov $\simeq$
245$''$$\times$155$''$). In all cases the colour code is green =
H$\alpha$, blue = {\OIII} and red = {\NII}. The images are displayed in a
logarithmic scale. North is up, east
to the left in each image.}\label{fig7}
\end{center}
\end{figure*}

\section{Final remarks}

Images obtained at sub-arcsec resolution (0.5$''$ -- 0.9$''$) have
revealed new morphological details in selected PNe. In IC\,351 and Vy\,1-1,
low-excitation, jet-like features have been 
detected for the first time. The morphology of these two objects is very
similar to this of well studied PNe like NGC\,6826 and NGC\,7009, strongly
suggesting that the same processes have operated in the formation of these
objects. In NGC\,6778 and K\,3-17 we found very complex structures. NGC\,6778
consists of a large number of cometary knots and filaments with two
systems of collimated outflows. K\,3-17 presents equatorial bubbles oriented at different
directions. Evidence for collimated outflows in K\,3-17 is 
provided by the spindle-like bipolar lobes. IC\,5217 and KjPn\,6 are bipolar
PNe consisting in a bright edge-on equatorial ring and faint bipolar lobes. A
comparison of the bipolar PNe in the sample
suggests that the formation processes of NGC\,6778, K\,3-17, and M\,2-48, implying the
action of collimated outflows, may differ from this of NGC\,650, K\,3-46 and BV\,1 which
fit the expectations of the GISW model.

\section*{Acknowledgments} 

This work has been supported partially by AYA2005-01495 of the Spanish MEC 
(co-funded with FEDER funds), 
and AYA2008-01934 of the Spanish MICINN (co-funded with FEDER founds), and 
by Consejer\'{\i}a de Innovaci\'on, 
Ciencia y Empresa of Junta de Andaluc\'{\i}a.



\begin{thebibliography}{}

\bibitem[Balick(1987)]{BAL87}
Balick, B. 1987, AJ, 94, 671

\bibitem[Balick et al.(1998)]{BAL98}
Balick, B., et al. 1998, AJ, 116, 360

\bibitem[Balick \& Frank(2002)]{BYF02}
Balick, B., \& Frank, A. 2002, ARA\&A, 40, 439

\bibitem[Bohigas(2001)]{BOH01}
Bohigas, J. 2001, Rev.~Mex.~A\&A, 37, 237

\bibitem[Bond, Pollacco, \& Webbink(2003)]{BPW03}
Bond, H.~E., Pollacco, D.~L., \& Webbink, R.~F. 2003, AJ, 125, 260 

\bibitem[Frank et al.(1993)]{FRA93}
Frank, A., Balick, B., Icke, V., \& Mellema, G. 1993, ApJ, 404, L25

\bibitem[Gieseking, Becker \& Solf(1985)]{GIE85}
Gieseking, F., Becker, I., \& Solf, J. 1985, ApJ, 295, L17 

\bibitem[Goncalves, Corradi, \& Mampaso(2001)]{DEN01}
Goncalves, D.~R., Corradi, R.,~L.~M. \&  Mampaso, A. 2001, ApJ, 547, 302

\bibitem[Guerrero, Chu, \& Miranda(2004)]{GUE04}
Guerrero, M.~A., Chu, Y.-H., \& Miranda, L.~F. 2004, AJ, 128, 1694

\bibitem[Guerrero et al.(2008)]{GUE08}
Guerrero, M.~A. et al. 2008, ApJ, 683, 272

\bibitem[Kaler et al.(1988)]{KAL88}
Kaler, J.~ B., Chu, Y.-H., \& Jacoby, G.~ H. 1988, AJ, 96,1407

\bibitem[Kwok, Purton, \& Fitzgerald(1978)]{KOW78}
Kwok, S., Purton, C.~R. \& Fitzgerald, P.~M. 1978, ApJ, 219, L125

\bibitem[L\'opez-Mart\'{\i}n et al.(2002)]{LOP02}
L\'opez-Mart\'{\i}n, L., et al. 2002, A\&A, 388, 652

\bibitem[Maestro, Guerrero, \& Miranda(2004)]{MGM04}
Maestro, V., Guerrero, M.~A., \& Miranda, L.~F. 2004, in ASP Conference Proceedings 313, 
Asymmetrical Planetary Nebulae III: Winds, Structure and the Thunderbird, ed. M. Meixner, 
J.H. Kastner, B. Balick \& N. Soker (San Francisco: Astronomical Society of the Pacific), p.127

\bibitem[Manchado et al.(1996)]{MAN96} 
Manchado, A., Guerrero, M.~A., Stanghellini, L., \& Serra-Ricart, M. 1996, The IAC 
Morphological Catalog of Northern Galactic Planetary Nebulae (La Laguna: Instituto de 
Astrof\'{\i}sica de Canarias)

\bibitem[Miranda et al.(2006)]{MIR06}
Miranda, L.~F., Ayala, S., V\'azquez, R., \& Guill\'en, P.~F.  2006, A\&A, 456, 591

\bibitem[Miranda et al.(2005)]{MIR05}
Miranda, L.~F., Ayala, S., Ulla, A., Olgu\'{\i}n, L., \& V\'azquez, R. 2005, in Planetary Nebulae as Astronomical 
Tools, Eds. R. Szczerba, G. Stasi\'nska, \& S.~K. G\'orny, IAP Conference Proceedings (New York), vol.\,804, 97

\bibitem[Miranda, Guerrero \& Torrelles(1999)]{MIR99}
Miranda, L.~F., Guerrero, M.~A., \& Torrelles, J.~M. 1999, AJ, 117, 1421

\bibitem[Miranda \& Solf(1992)]{MIR92}
Miranda, L.~F. \& Solf, J. 1992, A\&A, 260, 397

\bibitem[Miranda et al.(2001)]{MIR01}
Miranda, L.~F., Torreles, J.~M., Guerrero, M.~A., V\'azquez, R., \& G\'omez, Y. 2001, MNRAS, 321, 487

\bibitem[Montez et al.(2009)]{MON09}
Montez, R. Kastner, J.~H., Balick, B., \& Frank, A. 2009, ApJ, 694, 1491

\bibitem[Pereira et al.(2008)]{PER08}
Pereira, C.~B., Miranda, L.~F., Smith, V.~V., \& Cunha, K. 2008, A\&A, 477, 535

\bibitem[Sahai \& Trauger(1998)]{SAH98}
Sahai, R., \& Trauger, J.~T. 1998, AJ, 116, 1357

\bibitem[Schwarz, Corradi \& Melnick(1992)]{SCH92}
Schwarz, H.~E., Corradi, R.~L.~M., \& Melnick, J. 1992, A\&AS, 96, 23

\bibitem[Stanghellini, Corradi, \& Schwarz(1993)]{STA93}
Stanghellini, L., Corradi, R.~L.~M., \& Schwarz, H.~E. 1993, A\&A, 279, 521

\bibitem[V\'azquez et al.(2000)]{VAZ00}
V\'azquez, R., L\'opez-Mart\'{\i}n, L., Miranda, L.~F., Esteban, C.,
Torrelles, J. M., et al. 2000, A\&A, 357, 1031

\bibitem[V\'azquez et al.(1999)]{VAZ02}
V\'azquez, R., Miranda, L.~F., Torrelles, J.~M., et al. 2002, ApJ, 576, 860

\end{thebibliography}
\end{document}